\documentclass[aps,prd,preprint,nofootinbib]{revtex4}
\usepackage{amsfonts}
\usepackage{mathrsfs}
\usepackage{graphicx}
\usepackage{amsmath}
\usepackage{amssymb}
\usepackage{epsfig}
\usepackage{graphicx}
\newcommand{\bqa}{\begin{eqnarray}}
\newcommand{\eqa}{\end{eqnarray}}
\newcommand{\beq}{\begin{equation}}
\newcommand{\eeq}{\end{equation}}
\graphicspath{{fig/}{dia/}} \DeclareGraphicsExtensions{.eps}
\usepackage{subfig}
\begin{document}
\baselineskip 20pt
\title{NLO corrections to $\chi_{bJ}$ to two-$J/\psi$ exclusive decay processes\\}

\author{\vspace{1cm} Long-Bin Chen$^1$\footnote{
chenglogbin10@mails.ucas.ac.cn} and Cong-Feng
Qiao$^{1,2}$\footnote{qiaocf@ucas.ac.cn}\\}

\affiliation{$^1$School of Physics, University of Chinese Academy of
Sciences,  Yuquan Road 19A, Beijing 100049, China}
\affiliation{$^{2}$Collaborative Innovation Center for Particles and Interaction\\
USTC, Hefei 230026, China\\~\\}

\begin{abstract}
{~\\[-3mm]} The next-to-leading order QCD corrections
for $\chi_{bJ}$, the p-wave bottomonium, to $J/\psi$ pair decay
processes are evaluated utilizing NRQCD factorization formalism.
The scale dependence of $\chi_{b2}\rightarrow
J/\psi J/\psi$ process is depressed with NLO corrections, and hence
the uncertainties in the leading order results are greatly reduced.
The total branch ratios are found to be the order of $10^{-5}$ for
all three $\chi_{bJ}\rightarrow J/\psi J/\psi$
processes, indicating that they are observable in the LHC and super-B experiments.

\vspace {7mm} \noindent {PACS number(s): 12.38.Bx, 13.25.Gv,
14.40.Be }

\end{abstract}
\maketitle

\section{Introduction}

The advent of non-relativistic Quantum Chromodynamics (NRQCD) factorization formalism causes investigations on heavy quarkonium more reliable \cite{NRQCD}, which improves the understanding of strong interaction. It has been noted that for quarkonium production and decay, in many cases the leading order calculation in the framework of NRQCD is inadequate. The discrepancies between leading order calculations and experimental results are rectified by including higher order corrections, which has
encouraged various investigations. One typical example is the double
charmonium production in B-factory
\cite{belle,bralee,hqiao,lhch,ZYJ,GB1}. Furthermore, corrections of
higher order in the heavy-quark velocity $v$ are also important in
obtaining reliable predictions for quarkonium exclusive
decays \cite{fengjias,dfjia,lmchao}.

Inspired by the theoretical description of the double
charmonium production in B-factory, the bottomonium to double
charmonium decay processes have also been broadly investigated.
As in bottomonium to $J/\psi$ pair exclusive decay processes the
color-octet contribution is negligible, the theoretical evaluation
result is quite explicit in comparison with the experimental data.
The decays of $\eta_{b}$ to double $J/\psi$ \cite{Jia,GB2,Bra,SP},
$\Upsilon$ to $J/\psi+\chi_{cJ}$ \cite{jdf}, and $\chi_{bJ}$ to double $J/\psi$ \cite{FF,SWL,qiao} had been thoroughly
investigated. The process $\chi_{bJ} \rightarrow
J/\psi J/\psi(J=0,2)$ was evaluated in the frameworks of NRQCD and light cone distribution amplitude at the leading order in $\alpha_s$ expansion, except $\chi_{b1} \rightarrow J/\psi J/\psi$ process. In order to make the prediction more accurate and reduce the renormalization scale dependence, it is essential to evaluate the process with the NLO QCD corrections. In this aim, we calculate in the work the Next-to-Leading order(NLO) QCD corrections to the P-wave bottomonium $\chi_{bJ}$ decays to double $J/\psi$ in the framework of NRQCD. We obtain the completely analytical results for these three $\chi_{bJ} \rightarrow J/\psi J/\psi$ one loop processes.

The rest of this paper is organized as follows. Section II
presents the strategy and formalism for the study of
$\chi_{bJ}\rightarrow J/\psi J/\psi$ processes. Section III comprises numerical calculation and discussion on the results. Section IV gives  a summary and some conclusions. Several formulas are provided in the Appendix for reference.

\section{Calculation Scheme Description and Formalism}\label{II}

\begin{figure}[h]
\begin{center}
\includegraphics[scale=0.5]{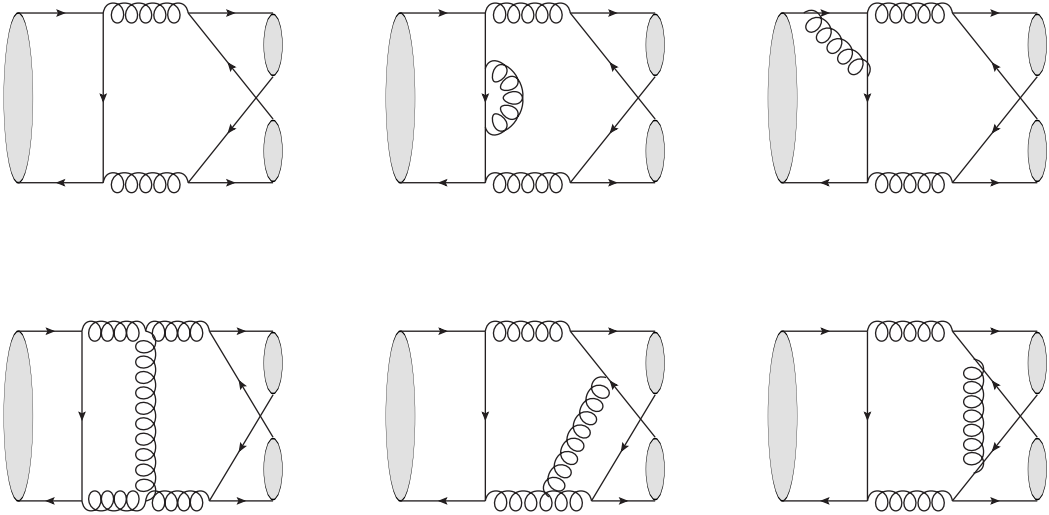}
\caption{Typical Feynman diagrams for $\chi_{bJ}\rightarrow J/\psi
J/\psi$ decays. \label{fig1}}
\end{center}
\end{figure}

In the calculation, we use \textbf{Mathematica} package \textbf{FeynArts} \cite{feynarts} to generate all the Feynman diagrams and decay amplitudes. Certain typical representative diagrams are shown in Fig.1. All the projection operators for quarkonium in this work may be found in
Ref.\cite{bodwpe,braaten}. For the spin-triplet color-singlet state,
the matrix product is expressed as:
\bqa v(\bar{p})\bar{u}(p)
=\frac{1}{4\sqrt{2}E(E+m)}(\not\!\bar{p}-m_c)
\not\!\epsilon_S^*(\not\!P+2E)(\not\!p+m_c)
\otimes\bigg(\frac{\bf{1}_c}{\sqrt{N_c}}\bigg)\ .
\label{projector}\eqa
Here $p=\frac{P}{2}+q,~\bar{p}=\frac{P}{2}-q$ respectively are the momenta of quark and antiquark, $\not\!\epsilon^*$ is a spin
polarization vector, $E^2=P^2/4=m_c^2-q^2$, $N_c=3$, and
$\bf{1}_c$ represents the unit color matrix. For the spin-singlet
and color-singlet state, the projection operator may be obtained by
replacing the $\epsilon_S^*$ in (\ref{projector}) with a $\gamma_5$.

Using the method described in \cite{braaten} and applying
\textbf{FeynCalc} \cite{Feyncalc} to the calculation of amplitudes,
the tree-level results are readily obtained. For One-loop QCD corrections,
more operations are necessary before arriving at the final result. For
$\chi_{b0 }$ and $\chi_{b2 }$ decays to the double $J/\psi$ processes,
there are 2 tree-level diagrams, 76 one-loop diagrams and 14 counter
terms. While for the $\chi_{b1}\rightarrow J/\psi J/\psi$ process, the contribution of gluon self-energy, triangle and four box diagrams
vanish, and there are only 44 one-loop diagrams functioning. The ultra
divergences are proportional to tree level amplitudes, hence
they vanish for $\chi_{b1}\rightarrow J/\psi J/\psi$ process
and hence no counter term is necessary.

In the calculation, Feynman diagrams and amplitudes are generated by \textbf{FeynArts}. \textbf{FeynCalc} is used to trace the matrices of spin and color, and to perform the derivation on the heavy quark relative momentum $q$ within quarkonium. The Mathematica package \textbf{Apart}
\cite{apart} reduces the propagator of each individual one-loop diagram. The
package \textbf{Fire} \cite{fire} is employed to reduce all one-loop
integrals to typical master-integrals. The calculation is performed in the Feynman gauge, while the Conventional
dimensional regularization with  $D=4-2\epsilon$ is adopted in regularizing the divergences. In the end, ultraviolet divergences are completely canceled by counterterms, the Coulomb
singularities are factored and attribute to the NRQCD long-distance matrix elements, and those infra divergences of short-distance coefficients cancel with each other, which confirms the NRQCD factorization for these processes at the NLO order.

In the expansion of strong coupling constant $\alpha_s$, decay width is formally expressed as:
\beq d\Gamma \propto |\mathcal {M}_{tree} + \mathcal
{M}_{oneloop}+\cdots|^2=|\mathcal {M}_{tree}|^2 + 2Re(\mathcal
{M}_{tree}^*\mathcal {M}_{oneloop}) + |\mathcal
{M}_{oneloop}|^2+\cdots\ . \nonumber \eeq
For $\chi_{b0,2}\rightarrow J/\psi J/\psi$ processes, up to the one-loop level, one only needs to keep the the first and second terms.
For $\chi_{b1}\rightarrow J/\psi J/\psi$ process, the tree level amplitude does not exist, and hence the leading order contribution comes from the one-loop amplitudes. By virtue of parity and Lorentz invariance, the decay amplitude for $\chi_{b1}\rightarrow J/\psi J/\psi$ can be constructed as following tensor structure:
\beq \mathcal
{M}(\lambda,\lambda_1,\lambda_2)=
(\varepsilon_{\mu\rho_1\rho_2\sigma}P_{\psi_1}^{\nu}-
\varepsilon_{\nu\rho_1\rho_2\sigma}P_{\psi_2}^{\mu})
\epsilon_{\psi_1}^{\mu} \epsilon_{\psi_2}^{\nu}P_{\psi_1}^{\rho_1}
P_{\psi_2}^{\rho_2}\epsilon_{\chi_b}^{\sigma}\mathcal {A}\ ,
\eeq
where $P_{\psi_1}$ and $P_{\psi_2}$ are the momenta of two final $J/\psi$s, $\epsilon$ denotes the polarization vector, and $\mathcal{A}$ is a scalar function of $m_b$ and $m_c$.

The ultraviolet and infrared divergences are contained in the renormalization constants $Z_2, Z_3, Z_m, Z_g$, correspond respectively to the quark field, gluon field, quark mass, and strong coupling constant $\alpha_s$. Among them, in our calculation the $Z_g$ and $Z_3$ are defined in the modified-minimal-subtraction $\overline{MS}$ scheme, while others are in the on-shell (OS) scheme. Thereafter, the counter terms read:
\begin{eqnarray}
\delta Z_2^{\rm OS}&=&-C_F\frac{\alpha_s}{4\pi}
\left[\frac{1}{\epsilon_{\rm UV}}+\frac{2}{\epsilon_{\rm IR}}
-3\gamma_E+3\ln\frac{4\pi\mu^2}{m^2}+4\right],
\nonumber\\
 \delta Z_m^{\rm OS}&=&-3C_F\frac{\alpha_s}{4\pi}
\left[\frac{1}{\epsilon_{\rm
UV}}-\gamma_E+\ln\frac{4\pi\mu^2}{m^2} +\frac{4}{3}\right], \nonumber\\
 \delta Z_3^{\overline{\rm
MS}}&=&\frac{\alpha_s}{4\pi} (\bar{\beta}_0-2C_A)
\left[\frac{1}{\epsilon_{\rm UV}} -\gamma_E + \ln(4\pi)\right], \nonumber\\
  \delta Z_g^{\overline{\rm MS}}&=&-\frac{\bar{\beta}_0}{2}\,
  \frac{\alpha_s}{4\pi}
  \left[\frac{1}{\epsilon_{\rm UV}} -\gamma_E + \ln(4\pi)
  \right].
\end{eqnarray}

Note that since the tree level contribution for $\chi_{b1}\rightarrow J/\psi J/\psi$ vanishes, the infrared divergences of short-distance coefficients appearing in the one-loop diagrams cancel with each other, and no ultraviolet divergences appear, this process is naturally finite in the next-to-leading order. The analytical results for the $\chi_{bJ}\rightarrow J/\psi J/\psi$ decay widths up to the one-loop level are given in the Appendix for reference.

\section{Results and discussion}

\subsection{Input parameters}

Before carrying out numerical calculation, the input parameters
need to be fixed. For the NRQCD matrix elements, those from
Refs. \cite{qiao,chao}, i.e. $\langle\mathcal
{O}_1\rangle_{\chi_{bJ}}=2.03\ \textrm{GeV}^5$ and $\langle\mathcal {O}_1\rangle_{J/\psi}=\frac{27 m_c^2 \Gamma(J/\psi\rightarrow
e^+e^-)}{8\pi\alpha^2(1-4C_F\alpha_s/\pi)} = 0.20\ \textrm{GeV} m_c^2$, are utilized. The charm quark and bottom quark masses are taken to be $m_c=1.5\pm0.1$ GeV and $m_b=4.9\pm0.1$ GeV respectively.
The masses of quarkonia are obtained from the PDG \cite{PDG}, which read
$M_{\chi_{b0}}=9.859$ GeV, $M_{\chi_{b1}}=9.892$ GeV, $M_{\chi_{b2}}=9.912$ GeV, $M_{J/\psi}=3.097$ GeV.
The two-loop expression for the running coupling constant $\alpha_s^l(\mu)$ with $n_l$ the number of light active flavors reads
\beq \frac{\alpha_s^{n_l}(\mu)}{4\pi}=\frac{1}{\beta_0 L}-\frac{\beta_1\ln
L}{\beta_0^3L^2}\ . \eeq
Here, $L=\ln(\mu^2/\Lambda_{QCD}^2)$,
$\beta_0=(11/3)C_A-(4/3)T_Fn_l$, and
$\beta_1=(34/3)C_A^2-4C_FT_Fn_l-(20/3)C_AT_Fn_l$, with
$\Lambda_{QCD}$ to be $339$ MeV and $n_l=3$, the number of light active flavors $(m_q\ll \mu)$. For numerical calculation in this work,
the renormalization scale is about the order of $m_c$, and
the strong coupling constant involves four active flavors, three light flavors and a massive flavor, the charm quark. Using a matching relation one may transit a running in terms of three active light flavors to a running in terms of four active flavors \cite{PDG}, i.e.,
\beq
\alpha_s^{n_l+1}(\mu)=\alpha_s^{n_l}(\mu)\bigg(1+\sum_{n=1}^{\infty} \sum_{i=0}^{n}c_{ni}[\alpha_s^{n_l}(\mu)]^n\ln^i(\frac{\mu^2}{m_c^2}) \bigg)\ .
\eeq
We take the two loop results for this expansion with the coefficients $c_{ni}$ are $c_{11}=\frac{1}{6\pi}, c_{10}=0, c_{22}=c_{11}^2, c_{21}=\frac{19}{24\pi^2}$, and $c_{20}=\frac{7}{24\pi^2}$ when $m_c$ is the pole mass. The $\alpha_s^{l+1}(\mu)$ is adopted in our numerical calculation and the $\bar{\beta}_0$ in counter terms $\{\delta Z_3, \delta Z_g\}$ is expressed as $\bar{\beta}_0=(11/3)C_A-(4/3)T_Fn_f$ with $n_f=3+1$.

\subsection{Results and Discussion}

After substituting the input parameters in the preceding subsection
to the analytical expressions, the numerical results are readily
obtained. The magnitudes of the decay widths for
$\chi_{bJ}\rightarrow J/\psi J/\psi$ are presented in Table~I, where the
first uncertainty comes from $m_b$ and the second one from the $m_c$. With the relation $\langle\mathcal {O}_1\rangle_{J/\psi} = 0.20\
\textrm{GeV} m_c^2$, the uncertainty of charmonium matrix element is attributed to that of $m_c$. In numerical evaluation, we let renormalization scale $\mu$ run from $m_b$ and $2m_b$ to estimate the uncertainties induced by even higher order contributions.
\begin{table}[h]
\caption{Decay widths for $\chi_{bJ}\rightarrow J/\psi J/\psi$ at
leading order and next-to-leading order, all in units of eV, with
$m_b=4.9\pm 0.1$ GeV, $m_c=1.5\pm 0.1$ GeV, and $\mu \in \{m_b, 2 m_b \}$.} %
\begin{center}
\renewcommand\arraystretch{1.4}
  \begin{tabular}{|l|c|c|c|c|c|}
    \toprule
    $\Gamma$(eV)& LO$\chi_{b0}$ & NLO$\chi_{b0}$ & LO$\chi_{b2}$
    & NLO$\chi_{b2}$ & $\textrm{LO} \chi_{b1}$\\
    \hline
    $\mu = m_b$& $7.66^{+1.33+0.75}_{-1.12-0.73}$ &
    $13.13^{+2.32+1.24}_{-1.93-1.18}$ &
    $19.54^{+3.94+3.55}_{-3.21-3.11}$ & $12.85^{+2.47+2.11}_{-2.03-1.90}$
    &
    $0.58^{+0.09+0.01}_{-0.08-0.01}$\\
    $\mu = 2m_b$& $3.56^{+0.59+0.35}_{-0.51-0.34}$
    & $7.85^{+1.34+0.76}_{-1.12-0.73}$ & $9.07^{+1.78+1.66}_{-1.45-1.45}$
    & $12.11^{+2.35+2.12}_{-1.93-1.88}$ & $0.18^{+0.03+0.00}_{-0.02-0.00}$\\
    \botrule
  \end{tabular}
\end{center}
\label{tab:kfac}
\end{table}

\begin{figure*}[hbtp]\centering
\subfloat[$\chi_{b0}\rightarrow J/\psi J/\psi$]{
  \begin{minipage}[t]{.5\linewidth}
    \includegraphics[width=\linewidth]{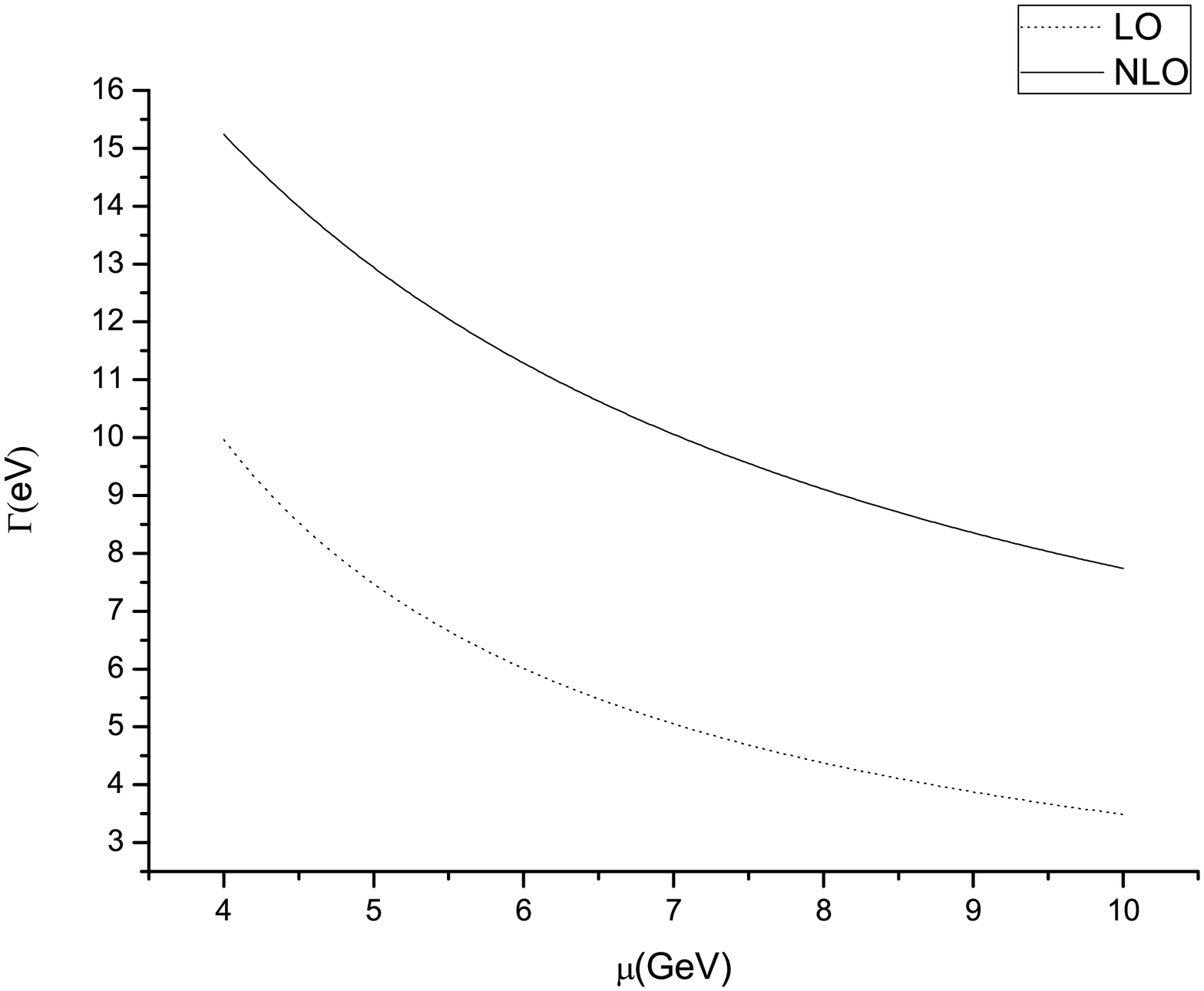}
  \end{minipage}
} \subfloat[$\chi_{b2}\rightarrow J/\psi J/\psi$]{
  \begin{minipage}[t]{.5\linewidth}
    \includegraphics[width=\linewidth]{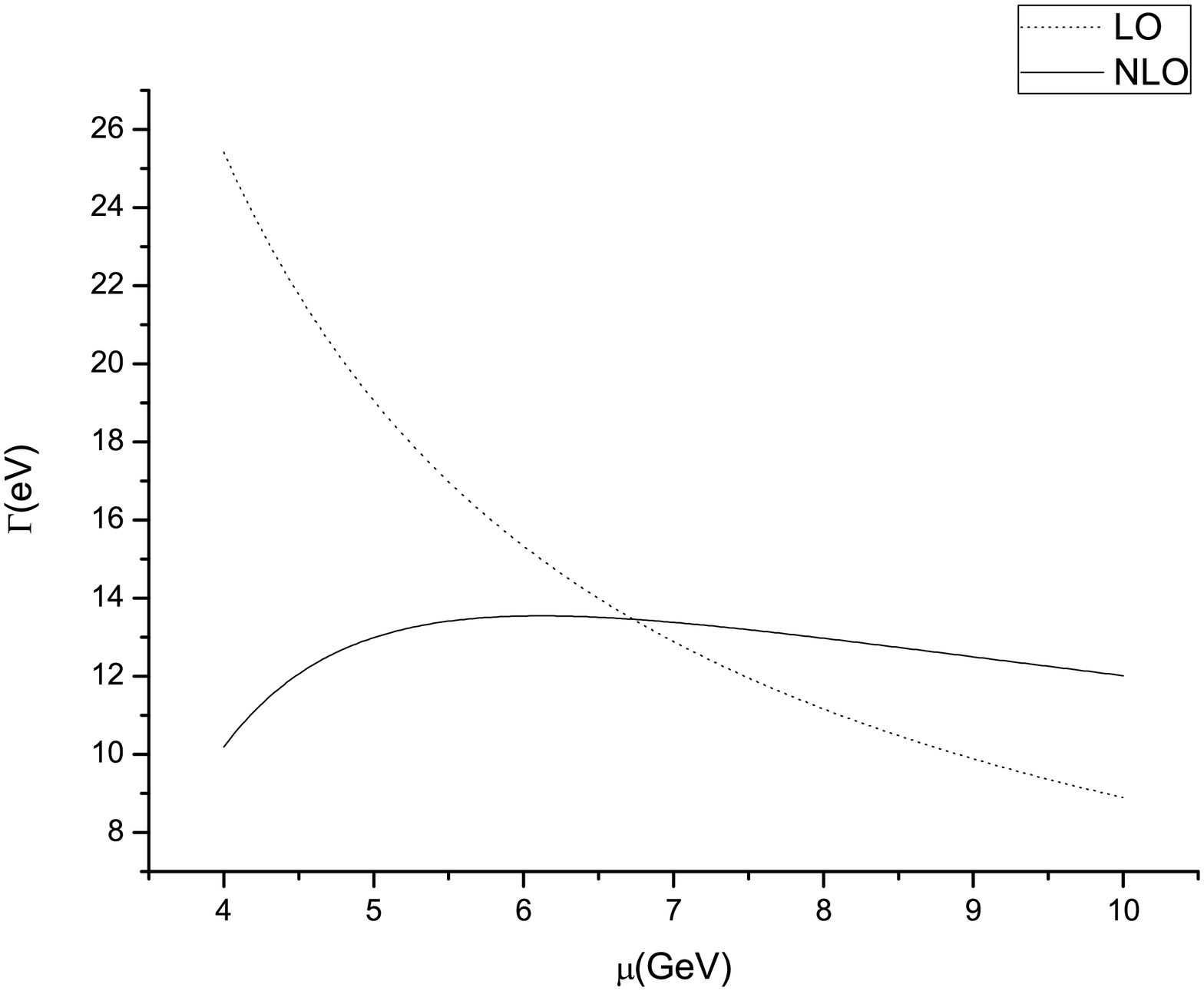}
  \end{minipage}
} \caption{\label{fig:mudep} The scale $\mu$ dependence for
$\chi_{bJ}\rightarrow J/\psi J/\psi(J=0,2)$ decay widths at LO and
NLO. Here $m_b=4.9\ \textrm{GeV}, m_c=1.5\ \textrm{GeV}, \textrm{and}\
 \Lambda_{QCD}=339\ \textrm{MeV}$.}
\end{figure*}

In Figure 2 we show the renormalization scale dependence of the leading order and the next-to-leading order decay widths of $\chi_{b0,2}\rightarrow J/\psi J/\psi$ processes. For these two processes, the decay widths are proportional to $\alpha_s^4(\mu)$ at the Born level, therefore having strong scale dependence. Fig. 2 exhibits that the $\mu$ dependence for $\chi_{b0}\rightarrow J/\psi J/\psi$ is still evident with the NLO QCD
corrections, while for $\chi_{b2}\rightarrow J/\psi J/\psi$ the $\mu$ dependence is substantially suppressed by the NLO QCD
corrections. Note that for $\chi_{b1}\rightarrow J/\psi J/\psi$ process, since the tree diagrams make null contribution, the effective leading order diagrams are at the one-loop level. So, in this work only the leading order
results are provided for the $\chi_{b1}\rightarrow J/\psi J/\psi$ decay process. Because the decay width of this process is proportional to $\alpha_s^6(\mu)$, the $\mu$ dependence is prominent in comparison with the other two processes. The decay width alters from 0.18 eV to 0.58 eV when the renormalization scale $\mu$ varying from $m_b$ to $2 m_b$.

\begin{table}[h]
\caption{\label{table2}%
Decay widths and branching ratios of $\chi_{bJ}\to J/\psi J/\psi$.
The first and second rows are the NLO results of this work with
scale $\mu=m_b$. The third and fourth rows are the leading order
results obtained in Refs. \cite{FF,SWL} with relativistic corrections.
The last row shows the branching ratio upper limits set by the Belle experiment \cite{belle1}.}
\begin{ruledtabular}
\begin{tabular}{lccc}
& $\chi_{b0}\to J/\psi J/\psi$  &$\chi_{b1}\to J/\psi
J/\psi$ &$\chi_{b2}\to J/\psi J/\psi$ \,\\
\hline
$\Gamma^{\text{NLO}}$(eV)                        &  $13.13^{+2.32+1.24+2.10}_{-1.93-1.18-5.39}$   &     $0.58^{+0.09+0.01+0.28}_{-0.12-0.01-0.40}$  &  $12.85^{+2.47+2.11+0.70}_{-2.03-1.90-2.66}$  \\
Br$^{\text{NLO}}$($10^{-5}$)                       &  $1.80^{+0.32+0.17+0.29}_{-0.26-0.16-0.74}$    &    $0.63^{+0.10+0.01+0.30}_{-0.13-0.01-0.43}$   & $5.85^{+1.12+0.96+0.32}_{-0.92-0.86-1.21}$  \\
$\Gamma^{\text{LO}}$(eV)~\cite{FF}              &   5.54   & $9.04\times10^{-7}$ & 10.6 \\
$\Gamma^{\text{LO}}$(eV)~\cite{SWL}             &   15   & $3.1\times10^{-4}$ & 35 \\
Br$^{\text{EXP}}$($10^{-5}$)~\cite{belle1}  &   $<$7.1 &   $<$2.7     &  $<$4.5   \\
\end{tabular}
\end{ruledtabular}
\end{table}

In Table II the theoretical predictions and experiment measurements for processes of p-wave bottomnium decays to $J/\psi$ pairs are given. Of our calculation, the next-to-leading order ones, there exist three main sources of uncertainties, i.e. from the uncertainties in charm quark mass, bottom quark mass and the variation of renormalization scale. Note that in practice the uncertainty in quarkonium non-perturbative matrix element also has a big effect in the calculation, whereas this effect is attributed to the charm quark mass through relation $\langle\mathcal {O}_1\rangle_{J/\psi} = 0.20
\textrm{GeV} m_c^2$ in our investigation as mentioned in above.

Once the Belle Collaboration had measured the $\chi_{bJ}\to J/\psi
J/\psi$ processes and obtained the corresponding upper limits for
branching ratios at the $90\%$ confidence level \cite{belle1},
which, as shown in the Table II, are compatible with our
estimations, i.e. at the order of $10^{-5}$. Though no significant
signals for $\chi_{bJ}\to J/\psi J/\psi$ have yet been observed at
the B-factory, from our calculation, they might be observed at the
LHC or future Super-B factory. According to Ref. \cite{bra}, while
$\sqrt{s}=14\text{TeV}$ the production cross sections of $\chi_{b0}$
and $\chi_{b2}$ at the LHC with are $\sigma(pp\rightarrow
\chi_{b0}+X)=1.5\mu b$ and $\sigma(pp\rightarrow \chi_{b2}+X)=2\mu
b$, respectively. In 2012, the luminosity of LHC was about
$50\text{fb}^{-1}$, so about $7.5\times 10^{10}\chi_{b0}$ and $
10^{11}\chi_{b2}$ were produced per year even at such luminosity.
This means that approximately thousands of $\chi_{b0,2}\rightarrow
J/\psi J/\psi\rightarrow l^+ l^- l^+ l^-$($l = e\ or\ \mu$)
processes happen each year at the LHC. As an example, the LHCb
detector covers a range of pseudo-rapidity $2<\eta<5$, and for the
dimuon event it requires the transverse momenta of the produced muon
pair satisfying $\sqrt{p_{T1}p_{T2}}>1.3$ GeV \cite{lhcb}.
Obviously, most of the events of the concerned processes satisfy
this requirement, hence the high detection efficency and the
substantial possibility of measuring the $\chi_{bJ}\rightarrow
J/\psi J/\psi$ processes at the LHC.


\section{Summary and Conclusions}

This work features a complete one-loop calculation for
$\chi_{bJ}\to J/\psi J/\psi$ exclusive decays in the framework of
NRQCD factorization. All infra divergences of short-distance coefficients are canceled out and hence confirms the NRQCD factorization in these processes at the next-to-leading order. Taking
$m_b=4.9\ \textrm{GeV}, m_c=1.5\ \textrm{GeV}$, numerical results show that the NLO QCD corrections for the decay width of $\chi_{b0}\to J/\psi J/\psi$ are all positive within the energy scale range of $m_b$ to $2 m_b$. While for $\chi_{b2}\to J/\psi J/\psi$, the NLO correction is negative at scale $\mu=m_b$ and positive at scale $\mu=2m_b$. As for $\chi_{b1}\to J/\psi J/\psi$ decay, the leading order process is at the one-loop level, which yields a notable number of events. Calculation results indicate that the renormalization scale $\mu$ dependence for $\chi_{b0}\to J/\psi J/\psi$ remains to be distinct with the NLO correction, while for $\chi_{b2}\to J/\psi J/\psi$ the $\mu$ dependence is substantially reduced. The branching ratios of all three $\chi_{bJ}\to J/\psi J/\psi$ decay processes are of the order $10^{-5}$. Although no evident signal has been observed in the
B-factory, with more high-luminosity and statistics, these p-wave bottomnium to double $J/\psi$ exclusive decay processes may be observed at the LHC or future Super-B experiment.


\vspace{0.07cm} {\bf Acknowledgments}

This work was supported in part by the National Natural Science
Foundation of China(NSFC) under the grant Nos. 10935012, 11121092, 11175249 and 11375200.

\vspace{1cm}

\appendix
\section{ amplitude}

The analytical results of the calculation are as follows. For the sake of simplicity of the expressions, we denote $a=\frac{m_c^2}{m_b^2}$. The leading order decay widths for $\chi_{b 0,2}\to J/\psi J/\psi$ are formulated as:
\beq
\Gamma_{\chi_{bJ}}^{LO}(J=0,2)=\frac{\pi^3\alpha^4_s(\mu)\langle\mathcal
{O}_1\rangle_{\chi_{bJ}}\langle\mathcal
{O}_1\rangle^2_{J/\psi}\mathbb{M}^J_{LO}}{(2J+1)373248m_b^7m_c^2
M_{\chi_{bJ}}}\sqrt{1-\frac{4M^2_{J/\psi}}{M^2_{\chi_{bJ}}}}
\eeq
with
$\mathbb{M}^0_{LO} = 1048576(1-4a+12a^2)/3$ and
$\mathbb{M}^2_{LO}  =  524288(13+56a+48a^2)/3$
for $\chi_{b0,b2}\to J/\psi J/\psi$ processes, and
\beq
\Gamma_{\chi_{b1}}^{LO}=\frac{\pi\alpha^6_s(\mu)\langle\mathcal
{O}_1\rangle_{\chi_{b1}}\langle\mathcal
{O}_1\rangle^2_{J/\psi}|\mathbb{M}^1_{LO}|^2}{34992m_b^{11}
M_{\chi_{b1}}(m_b^2-4m_c^2)^2}
\sqrt{1-\frac{4M^2_{J/\psi}}{M^2_{\chi_{b1}}}}
\eeq
for $\chi_{b1}\to J/\psi J/\psi$ process. Here,
\bqa \mathbb{M}^1_{LO}&&=
32m_b(42+212a-1421a^2+504a^3)
A_0(1)/(21a^2)+16m_b(-84+27a\nonumber\\&
&+4464a^2+712a^3)A_0(2)/(21a^2)-16m_b^3
(3515-1352a)B_0(1)/21\nonumber\\&
&+32m_b^3(1-28a-104a^2)
B_0(2)/(3a)-1728m_b^3(1-4a)B_0(4)/7\nonumber\\&
&+32m_b^3(1-7a+8a^2)B_0(5)/(3a)+16m_b^3
(84-111a-349a^2\nonumber\\&
&+2940a^3)B_0(6)/(21a^2)-32m_b^3
(2+15a+151a^2+72a^3)B_0(7)/(3a)\nonumber\\&
&+864m_b^5C_0(1)(1-4a)/7-32m_b^5C_0(2)
(3+10a)/3-64m_b^5C_0(3)(7\nonumber\\&
&+72a)/3-16m_b^5(815-126a-112a^2)C_0(4)/21
-32m_b^5C_0(5)(7-2a)/3\nonumber\\&
&+16m_b^5C_0(6)(45+288a-16a^2)/3-256m_b^5
C_0(7)(35+67a)/3\nonumber\\&
&+512m_b^5C_0(8)a(13-24a)/3-32m_b^5
(21-2365a+2213a^2-504a^3)/(21a)\ .\nonumber\\
\eqa

The NLO decay widths are formulated as :
\bqa \Gamma_{\chi_{bJ}}^{NLO}(J=0,2)
&=&\Gamma_{\chi_{bJ}}^{LO}\bigg(1 + \frac{16\alpha_s(\mu)
\textrm{Re}(131072\mathbb{M}^J_{NLO})}{\pi m_b^2\mathbb{M}^J_{LO}}
 \bigg).
\eqa
Here,
\bqa \mathbb{M}^0_{NLO}&&= A_0(1)(715-8959a+50042a^2-142196a^3+
169032a^4+26688a^5\nonumber\\& &-150912a^6)/(27a(1-9a+26a^2-24a^3))+
2A_0(2)(46-1015a+6578a^2\nonumber\\&
&-18023a^3+17406a^4+8688a^5-18720a^6)/(27a(1-9a+26a^2-24a^3))\nonumber\\&
&-((16n_l-177)-8(8n_l-91)a+ 12(16n_l-151)a^2)m_b^2B_0(1)/9+
(2\nonumber\\& &+31a-824a^2+2172a^3)m_b^2B_0(2)/(9a)-
11(1-4a+12a^2)m_b^2B_0(3)/3\nonumber\\&
&+2(11-11a+48a^2-180a^3)m_b^2B_0(4)/(3(1-3a))+ 2(1-18a+62a^2-152a^3\nonumber\\
&&+168a^4+864a^5-1536a^6)m_b^2B_0(5)/(9a(1-6a+8a^2))- 2(46-851a+5040a^2\nonumber\\
&&-11755a^3+6560a^4+4248a^5+17592a^6-29664a^7) m_b^2B_0(6)/(27a(1-9a\nonumber\\
&&+26a^2-24a^3))-2(2-11a-326a^2+738a^3+444a^4)m_b^2B_0(7)/(9a)\nonumber\\
&&-6a(11+6a)m_b^4C_0(1)-4a(19-52a+60a^2)m_b^4C_0(2)/9+8a(130a\nonumber\\
&&-79)m_b^4C_0(3)/9-2(20-207a+242a^2-24a^3)m_b^4C_0(4)/9-4(1+11a\nonumber\\
&&+12a^2-12a^3)m_b^4C_0(5)/9-8(15-89a+130a^2+54a^3)m_b^4C_0(6)/9\nonumber\\
&&+8(4-107a+102a^2)m_b^4C_0(7)/9+16(5-14a-8a^2+48a^3)m_b^4C_0(8)/9\nonumber\\
&&+8(1-4a+12a^2)m_b^2\ln(\frac{m_b^2}{\mu^2})/3
+ 32(1-4a+12a^2)m_b^2\ln(\frac{m_c^2}{\mu^2})/3\nonumber\\
&&+2(8n_l-561+(6693-88n_l)a
+2(208n_l-17631)a^2+ 6(15947-176n_l)a^3\nonumber\\
&&+48(24n_l-2383)a^4+23904a^5)m_b^2/(27(1-7a+12a^2)) \eqa
and
\bqa \mathbb{M}^2_{NLO}&&=
2A_0(1)(2041-9067a-16630a^2+96184a^3-7260a^4-133848a^5\nonumber\\&
&-45792a^6)/(81a(1-9a+26a^2-24a^3))-
4A_0(2)(655-3703a+13256a^2\nonumber\\&
&-64022a^3+154758a^4-38484a^5-163728a^6)/(405a(1-9a+26a^2-24a^3))\nonumber\\&
&-((520n_l-11097)+2(1120n_l-15527)a+
48(40n_l-379)a^2)m_b^2B_0(1)/135\nonumber\\&
&-2(5-201a-4702a^2-6096a^3)m_b^2B_0(2)/(135a)
-38(13+56a+48a^2)m_b^2B_0(3)/45\nonumber\\&
&+4(212+538a-1635a^2-3636a^3)m_b^2B_0(4)/(45(1-3a))-2(1+75a+142a^2+44a^3\nonumber\\
&&-776a^4-768a^5)m_b^2B_0(5)/(27a(1-2a))
+(524-1225a-13281a^2+48988a^3\nonumber\\
&&-26300a^4-7392a^5-24768a^6-55296a^7)m_b^2B_0(6)/(81a(1-9a+26a^2-24a^3))\nonumber\\
&&+2(2-1187a-2102a^2-2040a^3
-636a^4)m_b^2B_0(7)/(27a)\nonumber\\
&&+2(12-a-12a^2)m_b^4C_0(1)/3
+4(3+34a-82a^2-84a^3)m_b^4C_0(2)/27\nonumber\\
&&-8(3-133a-290a^2)m_b^4C_0(3)/27-2(4-61a-566a^2-360a^3)
m_b^4C_0(4)/27\nonumber\\
&&+8(2+16a+9a^2-6a^3)m_b^4C_0(5)/27-
2(396+1381a+572a^2-24a^3)m_b^4C_0(6)/27\nonumber\\
&&-4(275-820a+204a^2)m_b^4C_0(7)/27+
16(1+40a+32a^2-72a^3)m_b^4C_0(8)/27\nonumber\\
&&-8(13+56a+48a^2)m_b^2\ln(\frac{m_b^2}{\mu^2})/9+
16(13+56a+48a^2)m_b^2\ln(\frac{m_c^2}{\mu^2})/9\nonumber\\
&&+4(130n_l-3408-70(5n_l+126)a+(140163-1880n_l)a^2+
6(560n_l-21261)a^3\nonumber\\
&&+18(320n_l-21407)a^4-11880a^5)m_b^2/(405(1-7a+12a^2)) \eqa
with
\bqa
x_{1,2}=\frac{1\pm\sqrt{1-4a}}{2}\ ,\ y_{1,2}=\frac{1\pm\sqrt{1-4a}}{2\sqrt{a}}\ ,
\eqa
\bqa A_0(1)=A_0(m_c^2)=m_c^2(1+\ln(\frac{\mu^2}{m_c^2}))\ , \eqa \bqa
A_0(2)=A_0(m_b^2)=m_b^2(1+\ln(\frac{\mu^2}{m_b^2}))\ , \eqa \bqa
B_0(1)=B_0(m_b^2,0,0)=2+\ln(\frac{\mu^2}{m_b^2})+i\pi\ , \eqa \bqa
B_0(2)=B_0(4m_b^2,0,0)=2+\ln(\frac{\mu^2}{4m_b^2})+i\pi\ , \eqa \bqa
B_0(3)=B_0(m_b^2,0,m_b^2)=2+\ln(\frac{\mu^2}{m_b^2})\ , \eqa \bqa
B_0(4)=B_0(m_b^2,m_b^2,m_b^2)=
\ln(\frac{\mu^2}{m_b^2})+2-\frac{\pi}{\sqrt{3}}\ , \eqa \bqa
B_0(5)=B_0(m_b^2,m_c^2,m_c^2)=2+
\ln(\frac{\mu^2}{m_b^2})-\sum_{i=1}^{2}2x_i\ln(x_i)+i\pi\ , \eqa \bqa
B_0(6)=B_0(m_c^2,m_b^2,m_c^2)=2+ \ln(\frac{\mu^2}{m_c^2})-
\sum_{i=1}^{2}(x_i\ln(\frac{x_i-1}{x_i})-\ln(x_i-1))\ , \eqa \bqa
B_0(7)=B_0(2m_b^2+m_c^2,0,m_c^2)=2+
\ln(\frac{\mu^2}{m_c^2})-\frac{2}{2+a}(\ln(2a)-i\pi)\ , \eqa \bqa
C_0(1)=C_0(m_b^2,m_c^2,m_c^2,m_b^2,m_b^2,m_c^2)\ , \eqa \bqa
C_0(2)=C_0(m_b^2,m_c^2,2m_b^2+m_c^2,m_c^2,m_c^2,0)\ , \eqa \bqa
C_0(3)=C_0(4m_b^2,m_c^2,2m_b^2+m_c^2,0,0,m_c^2)\ , \eqa \bqa
C_0(4)=C_0(m_b^2,m_c^2,m_c^2,0,0,m_c^2)\ , \eqa \bqa
C_0(5)=C_0(m_b^2,m_c^2,m_c^2,m_c^2,m_c^2,m_b^2)\ , \eqa \bqa
C_0(6)=C_0(m_b^2,m_c^2,2m_b^2+m_c^2,0,0,m_c^2)\ , \eqa \bqa
C_0(7)=C_0(m_b^2,m_c^2,2m_b^2+m_c^2,0,m_b^2,m_c^2)\ , \eqa \bqa
C_0(8)=C_0(m_c^2,4m_c^2,2m_b^2+m_c^2,0,m_c^2,m_c^2)\ . \eqa

In above expressions, the divergent parts of the $A_0$ and $B_0$ functions have been removed, the value of $C_0$ function was given in \cite{SP} and we rechecked it analytically. The numerical evaluation was performed by means of the software LoopTools.


\end{document}